The Variation of the Solar Neutrino Fluxes over Time in the Homestake, GALLEX(GNO) and Super-Kamiokande Experiments

K. Sakurai[1,3], H. J. Haubold[2] and T. Shirai[1]
1. Institute of Physics, Kanagawa University, Yokohama 221-8686, Japan
2. Office for Outer Space Affairs, United Nations, P. O. Box 500, A-1400, Vienna, Austria
3. Advanced Research Institute for Science and Engineering, Waseda University, Shinjuku, Tokyo 169-8555, Japan

Abstract

  Using the records of the fluxes of solar neutrinos from the Homestake, GALLEX (GNO), and Super-Kamiokande experiments, their statistical analyses were performed to search for whether there existed a time variation of these fluxes. The results of the analysis for the three experiments indicate that these fluxes are varying quasi-biennially. This means that both efficiencies of the initial p-p and the PP-III reactions of the proton-proton chain are varying quasi-biennially together with a period of about 26 months. Since this time variation prospectively generated by these two reactions strongly suggests that the efficiency of the proton-proton chain as the main energy source of the Sun has a tendency to vary quasi-biennially due to some chaotic or non-linear process taking place inside the gravitationally stabilized solar fusion reactor. It should be, however, remarked that, at the present moment, we have no theoretical reasoning to resolve this mysterious result generally referred to as the quasi-biennial periodicity in the time variation of the fluxes of solar neutrinos. There is an urgent need to search for the reason why such a quasi-biennial periodicity is caused through some physical process as related to nuclear fusion deep inside the Sun.

1. Introduction

  Since the first attempt by R. Davis Jr. and his associates to detect the neutrinos from the Sun at the Homestake gold mine mine in the late 1960s (Davis et al., 1968), the observed records of the fluxes of solar neutrinos have been accumulated for the Homestake experiment for more than twenty years since then (Cleveland et al., 1998). During these years, other research groups have started their own experiments to observe the fluxes of these neutrinos: GALLEX (GNO), SAGE, Super-Kamiokande, and SNO experiments. In particular, both the GALLEX and SAGE experiments are substantially important, because they recorded neutrinos produced in the p-p reactions which are the first step of the proton-proton chain reactions of the main energy source of the Sun (see Table 1). All other three experiments, Homestake, SNO, and Super-Kamiokande, are only able to record the high energy neutrinos being produced from the PP-III process among the proton-proton chain reactions.

According to the newest observations available from the SNO experiment of the high energy neutrinos from the PP-III process, these neutrinos must be oscillating from one flavor to the other while traversing from the Sun to the Earth (Oser et al., 2004). Although the observed fluxes of these neutrinos are always deficient from the theoretical ones predicted from the standard model of the Sun, this deficit is essentially caused by the oscillatory change among three flavors of neutrinos, which is called the Mikheyev-Smirnov-Wolfenstein effect (Kim and Pevsner, 1993).

It should be, however, remarked that this deficit in the fluxes of solar neutrinos has been derived from averaging over all of the observed fluxes for many years. In doing so, the possible time variation of the fluxes of the solar neutrinos was neglected, because such variation could be thought of as a random fluctuation generated in the observing process of these fluxes. But, using the observed records available from the Homestake experiment, Sakurai (1979) took a view that this variation might have been correlated with some indices associated with the solar activity such as sunspot numbers. Then Sakurai tried to search for some possibility for the fluxes of the solar neutrinos to vary periodically. Based on such an idea, he analyzed the observed records and arrived at the conclusion that these fluxes have a tendency to vary quasi-biennially, though the phase as seen on the sunspot activity is delayed by about a year from the peak fluxes of the solar neutrinos.

This relation thus seemed to suggest that the process inside the core of the Sun was causally related to the photospheric phenomena as sunspot formation. Later, this quasi-biennial periodicity in the time variation of the fluxes of the solar neutrinos was named as the Sakurai periodicity (or Sakurai's periodicity) (Haubold, 1998).

According to the analysis recently made by Sakurai (2003), the observed records of the fluxes of the solar neutrinos obtained from the GALLEX (GNO) experiment also indicate that these neutrino fluxes seem to vary quasi-biennially, though it is impossible to conclude exclusively the existence of this periodicity, because these records have been obtained intermittently.

Recently, Shirai (2004) analyzed the observed records of the fluxes of the solar neutrinos obtained from the Super-Kamiokande experiment for about five years and has discovered that these fluxes has been varying with a period of about thirty months. This period is a little longer than the quasi-biennial one by a few months, but this difference is only within a statistical error range. So, it can be concluded that Shirai's result also supports the existence of the quasi-biennial periodicity in the time variation of the observed fluxes of the solar neutrinos in the Super-Kamiokande experiment.

Although the deficit as seen on the average fluxes of the solar neutrinos can be interpreted by referring to the basic nature of neutrinos as related to the

Mikheyev-Smirnov-Wolfenstein oscillation mechanism among three flavors of them, the cause for the time variation of the fluxes of the solar neutrinos must be looked for by taking into account some possible variability expected to exist in the efficiency of the proton-proton chain reactions. In this procedure, it is inevitable to modify the standard model of the Sun by incorporating processes that vary with time.

2. The Observed Records of Solar Neutrino Fluxes in the Homestake, GALLEX (GNO) and Super-Kamiokande Experiments

Since the early attempt by Davis et al. (1968) to observe the neutrinos from the Sun at the Homestake gold mine, Lead, South Dakota, the observed results of the fluxes of the solar neutrinos have been accumulated for more than two decades, which cover almost two solar activity cycles as shown in Fig. 1. These neutrinos were the ones which were identified as the final product from the PP-III process in the proton-proton chain reactions as summarized in Table 1 (Bahcall, 1989). The results shown in this figure seem to suggest that the fluxes of solar neutrinos tend to vary oscillatory over time. Taking this nature into considerations, Sakurai (1979) analyzed the data available from the Homestake experiment for the first five years to find out some possible periodic change in the observed fluxes of the solar neutrinos and reached a conclusion that there may exist the quasi-biennial periodicity in these fluxes.

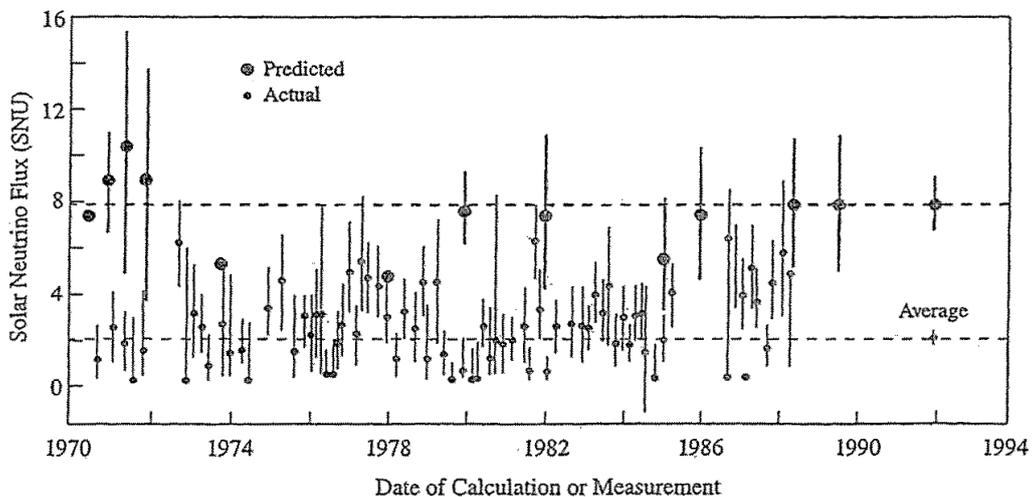

Fig. 1. Results of the observed fluxes of solar neutrinos from the Homestake experiment (Cleveland at al., 1998).

Since the Homestake experiment aimed to record neutrinos from the PP-III process in the proton-proton chain reactions, taking place in the core of the Sun, the observations of the neutrinos being produced from the first step of the proton- proton chain reactions must have been required to search for whether all steps of these chain reactions are varying quasi-biennially in the efficiencies. With respect to these steps, the most important is, of course, the first p-p reaction in the proton-proton chain reactions (Table 1). The experiments to record these neutrinos were initiated by using radiochemical techniques which used the reactions of the element Gallium with the low energy neutrinos from the p-p reactions (Kirsten, 2000).

Table 1.   The proton-proton chain reactions

| Reaction | step | Energy of neutrino ($\nu_e$) (MeV) |
|---|---|---|
| p-p reaction $p + p \rightarrow {}_1^2H + e^+ + \nu_e$ | 1a | $\geq 0.420$ |
| or $p + e^- + p \rightarrow {}_1^2H + \nu_e$ | 1b (*pep*) | 1.442 |
| PPI Process ${}_1^2H + p \rightarrow {}_1^2He + \gamma$ | 2 | |
| ${}_2^3He + {}_2^3He \rightarrow {}_2^4He + 2p$ | 3 | |
| PP III ${}_2^3He + {}_2^4He \rightarrow {}_4^7Be + \gamma$ | 4 | |
| ${}_4^7Be + e \rightarrow {}_2^7Li + \nu_e$ | 5 | (90%) 0.861 |
| ${}_3^7Li + p \rightarrow 2{}_2^4He$ | 6 | (10%) 0.383 |
| PP III process ${}_4^7Be + p \rightarrow {}_5^8B + \gamma$ | 7 | |
| ${}_5^8B \rightarrow {}_4^8Be + e^+ + \nu_e$ | 8 | < 15 |
| ${}_4^8Be \rightarrow 2{}_2^4He$ | 9 | |

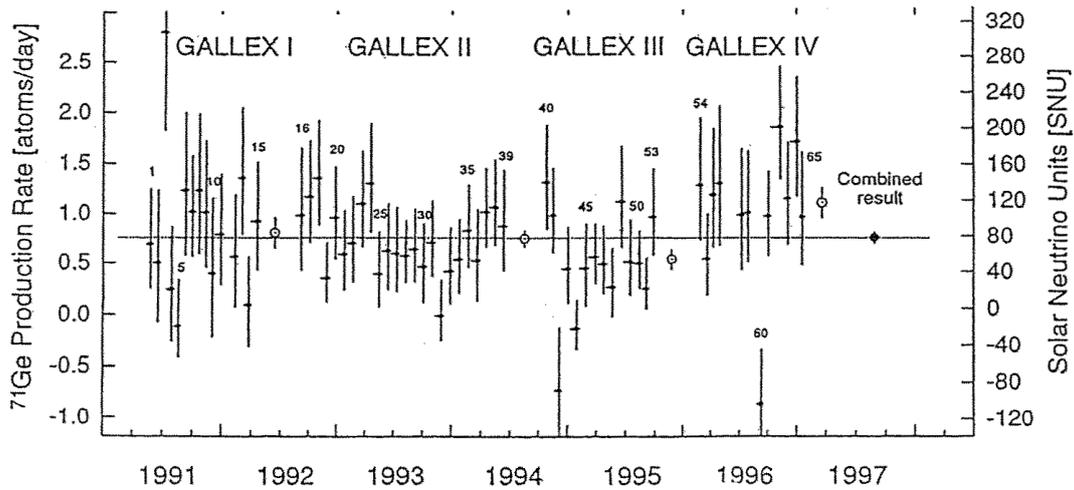

Fig. 2. Results of the individual runs of the GALLEX detector (Kirsten, 2000).

This experiment, being initially named as GALLEX, started in 1991 and has been continuing intermittently up to the present. The observed records of those neutrinos with the energy much lower than that of the neutrinos from the PP-III process are shown in Fig. 2. Despite that the SAGE experiment (Abdurashitov et al., 1996) was also initiated to detect the solar neutrinos produced from the p-p reactions at almost the same time as the GALLEX experiment, the observed results were not good in quality so that it was impossible to use them for our analysis to examine whether there exists the variation of those neutrino fluxes over time. Thus the records available from the SAGE experiment were not analyzed here in order to find out whether the data from the SAGE experiment indicate that the fluxes of the solar neutrinos vary with respect to time. Even if so, however, this experiment should be thought of as an important project to search for the time variation of the fluxes from the p-p reactions.

The solar neutrinos with relatively high energy from the PP-III process can be detected by observing the Cerenkov light that is emitted from electrons directly scattered by these neutrinos in the tank filled with pure water (Fukuda et al., 1996) or heavy water (Oser et al., 2004). The Super-Kamiokande experiment to detect such Cerenkov lights from neutrino interactions with electrons has accumulated the observed records long enough to analyze whether the observed fluxes of the solar neutrinos vary with time. These records have been compiled by Smy (1998) and plotted as shown in Fig. 3. They can be, therefore, used to look if there exists the time variation in the fluxes of these neutrinos.

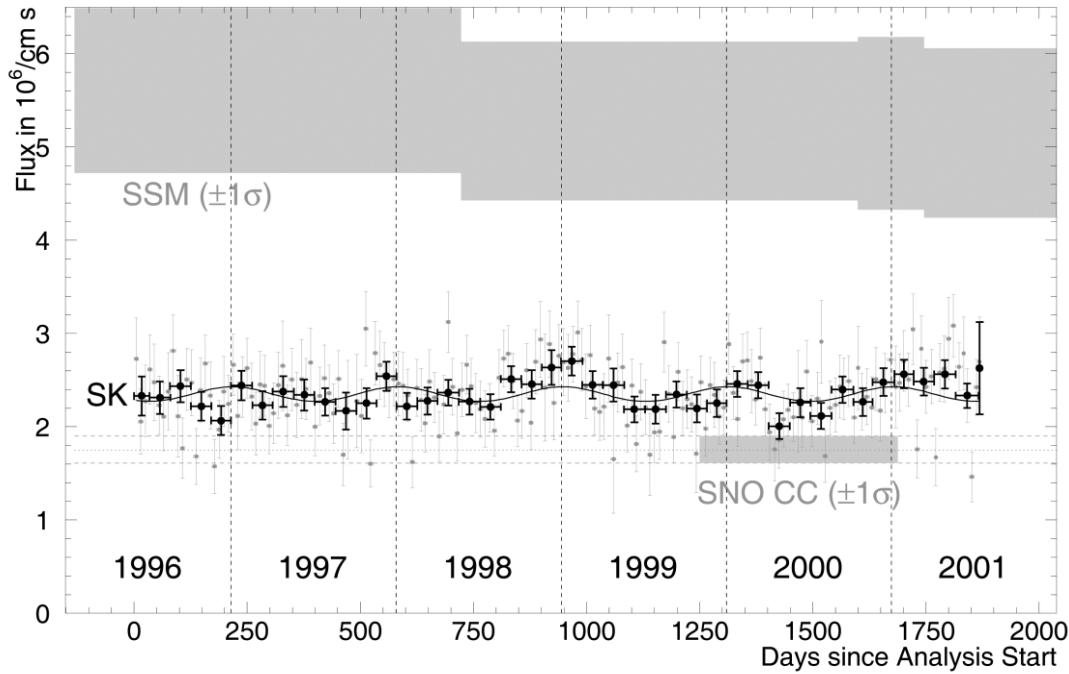

Fig. 3. Time variation of the fluxes of the solar neutrinos from the Super-Kamiokande experiments. The black points show the elastic scattering during 1.5 months. The gray points extend over a time period of 10 days. The solid line indicates the 7% flux variation expected from the change in the distance between the Earth and the Sun.

At the present moment, three observed records of the fluxes of solar neutrinos as mentioned earlier in this paper are ready for analysis to search for whether the fluxes of these neutrinos vary with time over a long-term record, though the observed records from the Super-Kamiokande experiment are still limited for several years. If we would be able to find out some evidence that the fluxes of solar neutrinos from both of the p-p reactions and the PP-III process have been varying cyclically over the observed period from about five years up to two decades, it would necessarily lead to conclude that the efficiency of the proton-proton chain reactions has a tendency to vary with some periodicity in time.

3. The Time Variation of the Solar Neutrino Fluxes as Discovered in the Analyses of the Observed Records

The possibility for the fluxes of solar neutrinos to vary with time was first proposed by Sakurai (1979) based on the analysis of the records from the Homestake experiment. According to his analysis, the fluxes of neutrinos must have been varying quasi-biennially with a period slightly longer than two years, e. g. about 26 months. He further predicted that the time variation of neutrinos might be causally associated with the quasi-biennial change in the sunspot activity seen as the deviation from the general trend of the eleven-year sunspot cycles.

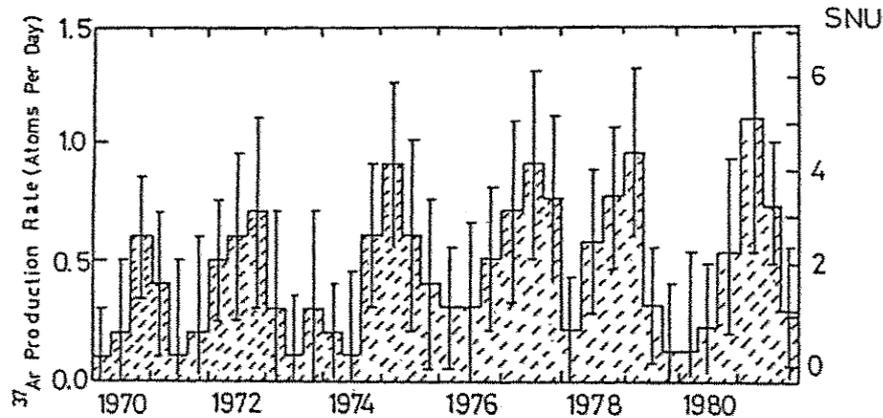

Fig. 4. The time variation of the fluxes of solar neutrinos for their four month averages. The original records are taken from the Homestake experiment (Sakurai, 1984).

Based on the analysis of the observed records up to the year 1994, which were available from the Homestake experiment, it was shown in Fig. 4 that observed fluxes of the solar neutrinos had a tendency to vary quasi-biennially, though it seemed that there existed the eleven-year change in these fluxes with small amplitude. Such an eleven-year change was once emphasized by Davis (1986) based on the comparison of the yearly variation of the fluxes of the solar neutrinos with that of the annual sunspot numbers.

According to Sakurai (1990), the observed records obtained by Davis and his associates (Davis, 1986; Cleveland et al., 1998) indicate that the time variation of the fluxes of solar neutrinos has a chaotic nature, since these fluxes tend to vary cyclically as shown in Fig. 5. The numbers in this figure are taken from the data points given by Davis (1986), which were plotted in the serial numbers in the acquisition of these fluxes. The period of each cyclic change as seen in this figure is almost equal to about 26 months, though this period was once extended to about 52 months, the doubling of the quasi-biennial period. This doubling occurred during the maximum period in the solar activity cycle between 1978 and 1983.

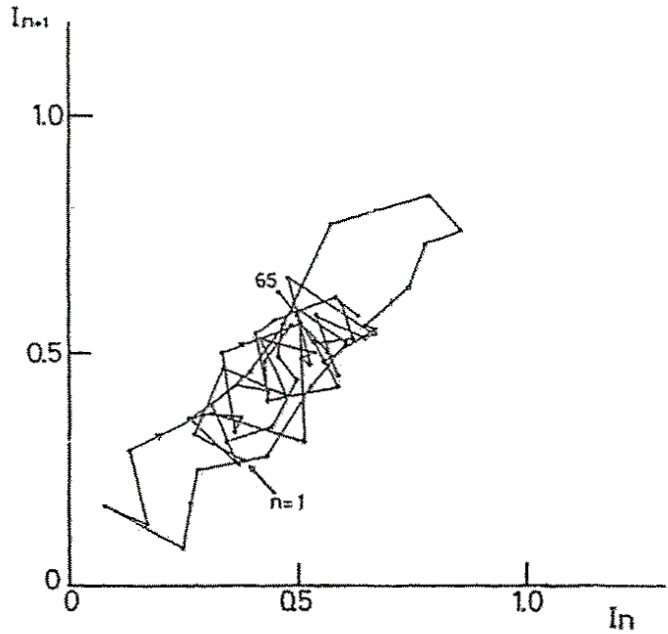

Fig. 5. A relation between two consecutive observed records defined as In and In+1 (n=1.2------64) for the Homestake experiment available from Davis (1986) and discussed by Sakurai (1990).

Both of the data from the Homestake and GALLEX (now GNO) experiments were analyzed since the latter began in 1991 (Kirsten, 2000). The observed records from the latter experiment were then statistically analyzed in the same way as done for the former one (Sakurai, 2003). The result of analyzed fluxes of solar neutrinos from the p-p reactions is highly suggestive as shown in Fig. 6 that these fluxes tend to vary quasi-biennially almost in phase with those obtained from the Homestake experiment. Fig. 6 further suggests that the efficiency of the neutrino production from the first step of the proton-proton chain reactions even has a tendency to vary quasi-biennially. Then, it may be natural to conclude that the time variation of the fluxes of solar neutrinos from the PP-III process is necessarily quasi-biennial, since this process is initiated by the p-p reactions as summarized in Table 1.

Recently, Shirai (2004) analyzed the observed records of the fluxes of solar neutrinos recorded by Super-Kamiokande experiment to discover whether these fluxes have a tendency to vary with time. In doing so, the seasonal components in these fluxes compiled by Smy (2001) were subtracted from the observed records to study the long-term variation of the production rate of the neutrinos from the PP-III process in the core of the Sun. The result obtained by Shiray (2004) is shown in Fig. 7. Although the observed period seems to be too short to deduce the periodicity in the variation of those fluxes over time, this periodicity is clearly seen in this figure and the period obtained is almost equal to 30 months. However, it can be said that this period is within the error range from a statistical point of view, though it is longer than 26 months. The result shown in Fig. 7 also supports that the time variation of the fluxes of the solar neutrinos observed at the Super-Kamiokande facility is taking place periodically with the quasi-biennial period.

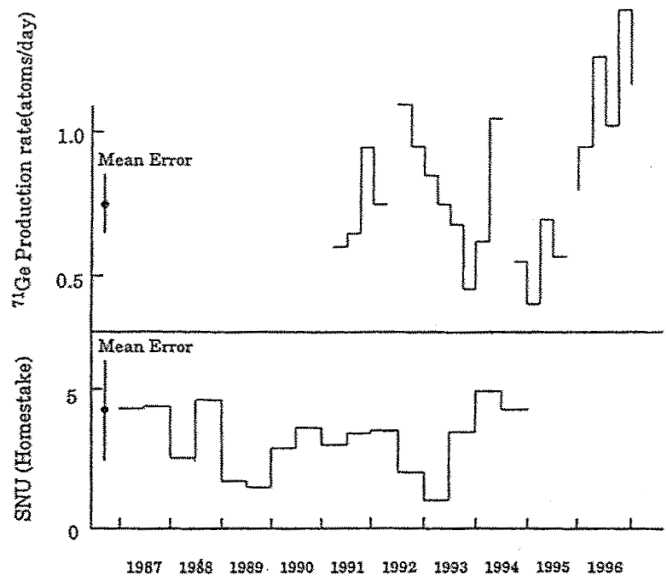

Fig. 6. The Time variation of the fluxes of solar neutrinos deduced from the GALLEX experiment (top) and the Homestake experiment (bottom) (Sakurai, 2003).

As shown in this paper, all of the observed records obtained from the Homestake, GALLX (GNO), and the Super-Kmiokande experiments show that the observed fluxes of the solar neutrinos vary quasi-biennially, though the amplitudes and the phases of their time variations are taking place in slightly different ways from each other. Since the records available from the GALLEX (GNO) experiment indicate that the efficiency of the p-p reactions, the first step of the proton-proton chain reactions, changes over time with a period of 26 months, it can be thought that this so called quasi-biennial variation of the fluxes of solar neutrinos is being produced due to a process in the solar core, which period should be quasi-biennial. It is thus of interest to look for a mechanism responsible for introducing the quasi-biennial periodicity as found in the production efficiency of the neutrinos associated with the proton-proton chain reactions taking place in the core of the Sun.

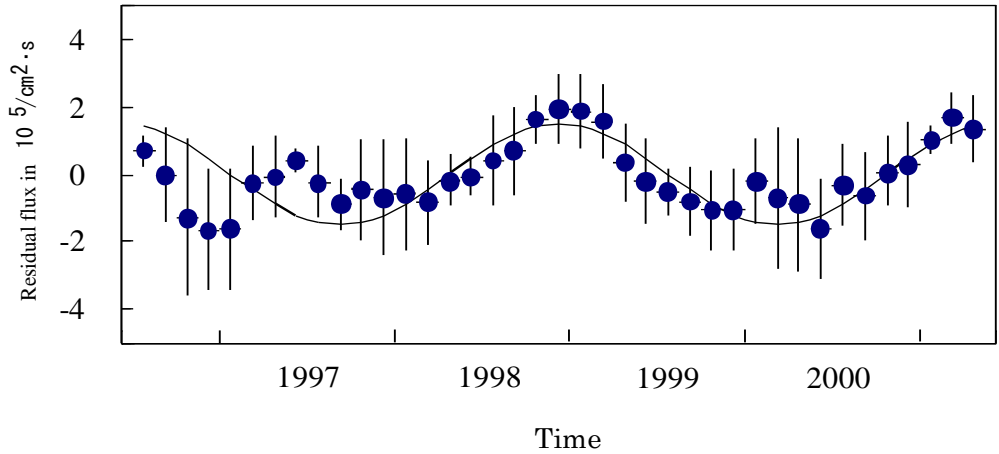

Fig. 7. The Time variation of the fluxes of the solar neutrinos obtained by the Super-Kmiokande experiments (Shirai, 2004).

4. Concluding Remarks

All of the observed records of the fluxes of the solar neutrinos available from the Homestake, GALLEX (GNO), and the Super-Kamiokande experiments indicate that the fluxes of solar neutrinos being produced from both of the p-p reactions and the PP-III process in the proton-proton chain reactions are varying quasi-biennially. Since this quasi-biennial periodicity is found in the neutrino production in the p-p reactions, the origin of this periodicity must be connected with some periodic change in the internal physical processes that initiate these reactions in the core of the Sun.

At the present moment, there is no conclusive interpretation to resolve the origin of the quasi-biennial periodicity in the time variation of the fluxes of solar neutrinos, but it might be said that there exists some unknown synchronous mechanism for the internal physical processes which are responsible for the quasi-biennial periodicity in the efficiency of the neutrino production in the proton-proton chain reactions inside the core of the Sun. This mechanism seems to be causally connected with some non-linear processes that can produce the cyclic change in the fluxes of solar neutrinos as shown in Fig. 5.

Acknowledgement: The authors are quite indebted to Raymond Davis, Jr. (1914-2006) and many other colleagues for their criticisms and suggestions during the course of the studies described in this paper.